\documentclass[12pt,preprint]{aastex}
\usepackage{epsfig}
\usepackage{natbib,lscape}
\usepackage{graphicx,color}
\newcommand\etal{{\it et al.}}
\def\red#1 {\textcolor{red}{#1}~}
\def\grn#1 {\textcolor{green}{#1}~}
\def\blue#1 {\textcolor{blue}{#1}~}

\shorttitle{High Obscuration in H$_2$O Maser AGN}
\shortauthors{Greenhill, Tilak \& Madejski}

\begin{document}
\title{Prevalence of High X-ray Obscuring Columns  among AGN that Host H$_2$O  Masers}

\author{Lincoln J. Greenhill, Avanti Tilak}\affil{Harvard-Smithsonian Center for Astrophysics, 60 Garden Street, Cambridge, MA 02138}\email{greenhill@cfa.harvard.edu}
\and
\author{Grzegorz Madejski}\affil{Stanford Linear Accelerator Center and Kavli Institute for Astrophysics and Cosmology, Menlo Park, CA 94025}

\begin{abstract}

Of 104 AGN known to exhibit H$_2$O maser emission, X-ray data that enable estimation of column densities, or lower limits, are available for 42.  Contributing to this, we report analysis of new and archival X-ray data for 8 galaxies and collation of values for  three more.  Maser emission is indicative of large columns of cold gas, and in five of the eight new  cases, maser spectra point toward origins in accretion disks viewed close-to edge-on (a.k.a. ``disk-maser'' systems).  In these, we detect hard continuum and Fe K$\alpha$ emission with equivalent widths on the order of 1\,keV, which is consistent with Compton reflection, fluorescence by cold material, and obscuring columns $\ga 10^{24}$\,cm$^{-2}$.    Reviewing the full sample of 42, 95\% exhibit N$_{\rm H} >10^{23}$\,cm$^{-2}$ and 60\% exhibit N$_{\rm H} >10^{24}$\,cm$^{-2}$.  Half of these are now recognized to be disk masers (up from 13); in this sub-sample, which is likely to be more homogeneous vis-\'a-vis the origin of maser emission, 76\% exhibit N$_{\rm H} >10^{24}$\,cm$^{-2}$.  The probability of a common parent distribution of columns for disk-masers and other AGN masers is $\la 3\%$.   Because ground-based surveys of AGN to detect new disk masers are relatively unbiased with respect to X-ray brightness and comparatively inexpensive, they may also be efficient guides for the sensitive pointed X-ray observations required to identify Compton-thick objects outside of shallow surveys. 

\end{abstract}


\section{Introduction}

Extragalactic H$_2$O maser emission at 22 GHz is known in 104 active galactic nuclei (AGN)(\cite{zha06}, \cite{kon07}, \cite{braatz08}, Braatz et al. 2008, in preparation). In early analyses, the majority of the nuclei for which pointed X-ray observations exist  are Compton thick or nearly so, with columns (N$_{\rm H}$) $\ga 10^{24}$ cm$^{-2}$ \citep{mad06, zha06, gree07}. \citet{zha06} consider  all maser-host galaxies, where the emission is associated with either nuclear activity or star formation. In a sub-sample of 26 with apparent maser luminosity $>10$ L$_\odot$ (a.k.a. ``megamasers''),   85\% exhibited N$_{\rm H}$ $>10^{23}$ cm$^{-2}$ and 50\% exceeded $10^{24}$ cm$^{-2}$.  \citet{mad06} focus on a probably more homogeneous sub-sample for which maser spectra suggest the emission specifically traces rotating, highly inclined disk structures close to the central engines.  (We refer to these as ``disk masers.'') Seven of 9 nuclei with disk masers were Compton thick.

It is plausible that AGN which host masers are characterized by high N$_{\rm H}$  in general, because large reservoirs of molecular gas along the line of sight are required to produce substantial maser amplification \citep{eli82}.  Among disk masers, prevalence of high N$_{\rm H}$ may be anticipated because  the emission is believed to arise in inclined disk structures.   We have analyzed new, archival, and published X-ray data for maser AGN and examine the statistics of the updated sample, now about one third larger than that of \citet{zha06}.  Masers show promise as signposts of Compton-thick AGN that are readily detected from the ground.  In $\S2$, we describe new spectral analyses and estimation of N$_{\rm H}$.  We discuss selection effects in the full sample and inferences about maser AGN in $\S3$.

\section{Data Analysis}
We observed seven maser-host AGN using the XMM-Newton observatory. In addition, we identified  unpublished observations of five  disk-maser hosts in the XMM and Chandra archives (12 targets total).  All data were reduced using standard tasks in SAS and CIAO.  XMM observations of NGC 6323 and NGC 6264 were lost due to flaring; NGC 3735 and NGC 2639 were lost due to high background. Table\,1 lists the final sample.

XMM EPIC PN spectra were extracted from 30$\arcsec$ circles centered on the brightest pixels within 2$\arcsec$ of the target coordinates (comparable to the 1.5$\arcsec$ astrometric accuracy). Background spectra were extracted from 80$\arcsec$ circles, centered off-source, as far from the readout nodes as the targets.  We did not use EPIC MOS data due to smaller effective area above 2\,keV.  Chandra ACIS target spectra were extracted from 6$\arcsec$ circles, while the backgrounds were obtained from annuli centered on the targets, with radii of 15$\arcsec$ and 30$\arcsec$. The spectra were grouped with 10 counts bin$^{-1}$  and analyzed using XSPEC11. 

\begin{deluxetable}{lccccc}
\tablecolumns{6}
\tablewidth{4.2in}
\tablecaption{Observation Summary}
\tablehead{
\colhead{Galaxy}  & 
\colhead{Tel.} &
\colhead{Epoch\tablenotemark{(a)} }&
\colhead{T\tablenotemark{(b)} }&
\colhead{Counts} &
\colhead{ Ref\tablenotemark{(c)} }
}
\startdata
NGC\,1320  &  XMM &   06-08-06 & 12260 & 2778 & 1\\
NGC\,1194 &  XMM &   06-02-19 & 12606 & 1083 & 2 \\
NGC\,6926  &  XMM &   05-11-19 &   8875 & 149 & 3 \\
Mrk\,1419   &  XMM &   05-10-30 & 22298 & 518 & 4 \\
Mrk\,34    & XMM  &   05-04-04 &   8905 & 682 & 5 \\
NGC\,5793  & XMM  &  05-07-21 &  15686 & 388 & 6 \\
ESO\,013-G012 & XMM & 03-04-21 &  21625 & 167 & -- \\
VV\,340A     & Chandra &  06-12-17 &  14862 & 125 & 7 \\
\vspace{-0.1in}
\enddata
\tablenotetext{(a)}{Observing date in  {\it yymmdd} format. }

\tablenotetext{(b)}{Duration of calibrated data used in modeling, in seconds.}

\tablenotetext{(c)} {References for disk-maser identification in targets:  (1) \citet{kon07}, (2) (Braatz \etal, 2008, in preparation), (3) \citet{greenhill03}, (4) \citet{kon06}, (5) \citet{kon06}, (6) \citet{hag97}, (7) \citet{kon06}.   \vspace{-0.05in} \\ } 
\label{tablog}
\end{deluxetable}

\begin{figure}[th]
\begin{center}
\scalebox{1}[0.95]{\plotone{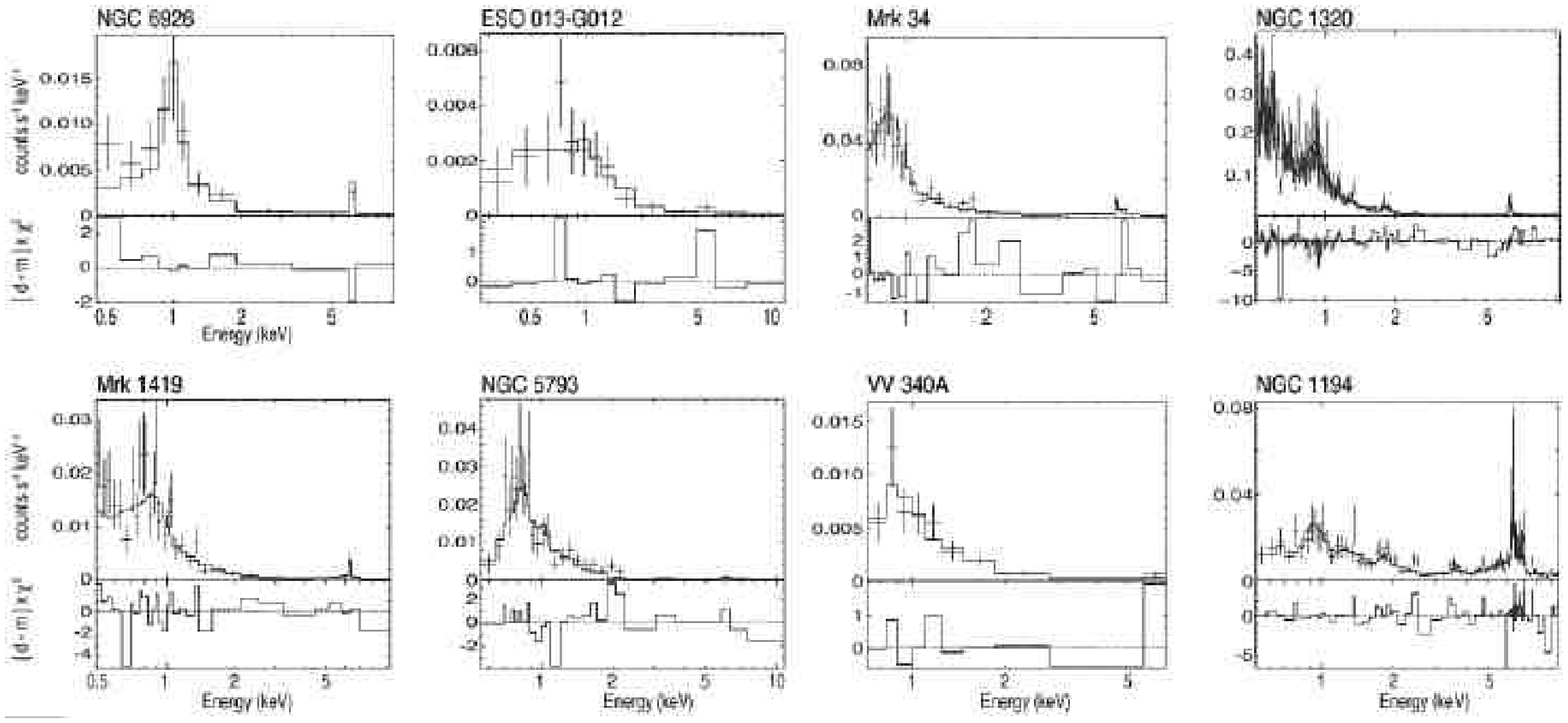}}
\caption{XMM EPIC PN spectra of 8 AGN that host H$_2$O masers. Each source was fitted with thermal emission model and a Compton reflection component (Tab.\,2).  All except VV\,340A and ESO\,013-G012 exhibit Fe K$\alpha$ emission at 6.4\,keV (rest). Mrk\,34, NGC\,1320, and NGC\,1194 also show Fe K$\beta$ at 7.05\,keV (rest).  NGC\,1194 exhibits evidence as well for (1) direct transmission of the central engine power law spectrum, obscured below $\sim 5$\,keV by a 1.06$_{-0.24}^{+0.36} \times 10^{24}$ cm$^{-2}$ column and (2) K$\alpha$ emission from Si, S, and Ca.}  
\label{figspectra}
\end{center}
\vspace{-0.2in}
\end{figure}

All 8 observations are characterized by relatively hard continua above 2\,keV.   Importantly, this is accompanied in five sources by Fe K$\alpha$ emission with large equivalent widths (EWs), regardless of continuum model (Tab.\,2;  Fig.\,\ref{figspectra}).   The strong lines probably arise from continuum reprocessing via fluorescence when the central engine is detected only in Compton reflection \citep[e.g.,][2006]{lev02}.  Since signal to noise is relatively low for a majority of targets, we adopted a ``minimum'' model with a common set of fixed and free parameters.    Because of degeneracies, results are best viewed as indicative.   We assumed that the continuum is entirely described by reflection (a PEXRAV model) with underlying (incident) photon index fixed to 1.7. Larger values \citep[$\Gamma\sim 2$; e.g.,][]{zdz96} observed toward  type-1 AGN are also consistent.  We modeled the Fe K$\alpha$ lines as Gaussians with fixed 6.4\,keV rest-frame energy and $\sigma=20$\,eV intrinsic width.   We also added a MEKAL component to match better emission below 2\,keV. Fitted sub-solar  abundances, when these arose, were probably not an  accurate reflection of composition, but rather a consequence of a one-temperature model and omission of  photoionization effects.  Separate absorption columns were applied to MEKAL and PEXRAV plus line components. 

Spectral fitting with even the minimum model shows Fe K$\alpha$ line emission in Mrk\,34, Mrk\,1419 and NGC\,6926, as well as marginally in NGC\,5793 (Tab.\,2; Fig.\,\ref{figspectra}).  We compute EWs with respect to the reflected continuum.  In the case of NGC\,5793, we obtained EW=2.38\,keV($\chi^2$/d.o.f=26.6/26) with a 0\,keV bound.  Freeing the line energy, we obtained EW=2.46$_{-2.25}^{+2.67}$\,keV and $\chi^2$/d.o.f.= 24.9/24, for line energy 6.32$_{-0.42}^{+0.09}$\,keV (rest).  The shift in line energy may be due to coarse binning.  Some elaboration on the minimum model was required for Mrk\,34, to accommodate excess near 7\,keV, which we fit  with a second Gaussian at 7.05\,keV (rest) with $\sigma=20$\,eV, corresponding to Fe K$\beta$. EWs $\ga 1$\,keV  for Mrk\,34, Mrk\,1419, and NGC\,6926 are consistent with N$_{\rm H}\ga$ 1-2$\times$10$^{24}$ cm$^{-2}$ \citep{lev06}, though quantitative interpretation depends on geometry.  This and intrinsic limitations of the minimum model argue for inferred columns to be best regarded as indicative. No estimates are possible for ESO\,013-G012 and VV340A because neither exhibited clear Fe K$\alpha$ emission (though both showed marginal excess at $\sim 6$\,keV).

\begin{deluxetable}{lccccccrr}
\rotate
\label{ }
\tablecolumns{9}
\tablewidth{8in}
\tablecaption{Summary of Spectral fitting}
\tablehead{
\colhead{Source} &
\colhead{n$_{\rm H}$\tablenotemark{(a,b)} } &
\colhead{$\Gamma$\tablenotemark{(b)} } &
\colhead{kT} &
\colhead{$Z_{\sun}$/Z \tablenotemark{(b)} } &
\colhead{Fe K$\alpha$ EW\tablenotemark{(c)} } &
\colhead{n$_{\rm H}$\tablenotemark{(d)} } &
\colhead{$\chi^2$/d.o.f.} &
\colhead{L$_{2-10\,{\rm keV}}$\tablenotemark{(e)} }  \\
\colhead{} & 
\colhead{(10$^{20}$ cm$^{-2}$)} &
\colhead{} & 
\colhead{(keV)} &
\colhead{} & 
\colhead{(keV)}&
\colhead{(10$^{24}$ cm$^{-2}$)} &
\colhead{} & 
\colhead{(ergs s$^{-1}$)}
}

\startdata
NGC\,1320\tablenotemark{(f)}& 65$_{-5}^{+6}$& 1.42$_{-0.48}^{+0.55}$& 0.66$_{-0.05}^{+0.11}$& [1.0]& 2.20$_{-0.43}^{+0.44}$& $\ga 1$ & 133/134& 7.5$\times$10$^{41}$\\

NGC\,1194\tablenotemark{(g)}& 99$_{-73}^{+32}$& 2.58$_{-0.44}^{+0.64}$& 0.91$_{-0.26}^{+0.31}$& [1.0]& 2.11$_{-0.50}^{+0.50}$&  1.06$_{-0.24}^{+0.36}$ &69/74 & 3.7$\times$10$^{41}$\\ 

NGC\,6926& [7.43]& [1.7]& 1.12$_{-0.21}^{+0.55}$& [1.0]& 3.38$_{-3.21}^{+0.71}$& $\ga 1$ &7.6/8 & 5.8$\times$10$^{40}$\\

Mrk\,1419 & [3.8] & [1.7] & 0.76$_{-0.11}^{+0.12}$& [0.05]& 1.76$_{-0.88}^{+0.89}$& $\ga 1$ & 35/30& 3.2$\times$10$^{40}$\\

Mrk\,34\tablenotemark{(h)} & [0.69] & [1.7] & 0.79$_{-0.10}^{+0.09}$ & [0.20]& 1.14$_{-0.59}^{+0.61}$ & $\ga 1$ & 27/25& 9.8$\times$10$^{41}$\\

NGC\,5793& 39$_{-16.3}^{+12.8}$& [1.7]& 0.48$_{-0.10}^{+0.11}$& [0.1]& 2.46$_{-2.25}^{+2.67}$  & -- & 25/24 & 1.8$\times$10$^{41}$\\

ESO\,013-G012& [7.39]& [1.7]& 1.43$_{-0.51}^{+2.13}$& [0.1]& -- & -- & 5.7/11& 1.1$\times$10$^{40}$ \\

VV\,340A & 90$^{+28}_{-24}$ & [1.7]& 0.62$^{+0.18}_{-0.25}$& [1.0]& -- & -- & 5.1/6& 4.9$\times$10$^{41}$ \\
\enddata
\small
\tablenotetext{(a)}{
Absorption due to Galactic and host-galaxy interstellar media.}

\tablenotetext{(b)}{
Values held fixed in fitting are enclosed in brackets.}

\tablenotetext{(c)}{
Computed with respect to the reflected continuum and for fixed 6.4\,keV line energy (rest), except for the marginal detection in NGC\,5793. }

\tablenotetext{(d)}{Absorption within the AGN.  Inferred indirectly from large Fe K$\alpha$ EWs or directly  from the absorbed power law component of NGC\,1194.}

\tablenotetext{(e)}{Apparent luminosity other than for NGC\,1194, where intrinsic luminosity is  $4.4\times10^{42}$\,ergs\,s$^{-1}$. }

\tablenotetext{(f)}{
Model includes a second MEKAL component with fixed temperature of 0.08\,keV and a Gaussian component corresponding to Fe K$\beta$ emission, fixed at 7.05\,keV (rest) with intrinsic width $\sigma=20$\,eV and fitted EW of 282$_{-56}^{+55}$eV.  Normalization for the Fe K$\beta$ line was constrained to be $0.11\times$ the normalization for the Fe K$\alpha$ line \citep{geo91}.} 

\tablenotetext{(g)}{
As in (f) regarding secondary MEKAL and Fe K$\beta$ components (EW=299$_{-71}^{+70}$ eV).  Additional Gaussian components representing blends of ionization states were fitted with observed energies of $1.75\pm0.10$\,keV (Si K$\alpha$; EW=160$_{-160}^{+240}$), $2.49\pm0.06$\,keV (S K$\alpha$; EW=390$_{-260}^{+340}$), and $3.58\pm0.13$\,keV (Ca K$\alpha$; EW=380$_{-240}^{+230}$). The photon indices for incident reflected and transmitted components are set equal.}

\tablenotetext{(h)}{
As in (f) regarding addition of an Fe K$\beta$ component (EW=150$_{-90}^{+0.70}$ eV). \vspace{-0.2in} } 
\end{deluxetable}

Comparatively high signal-to-noise ratios enabled fitting more complex models to data for NGC\,1194 and NGC\,1320.  For NGC\,1320, we were able to fit a photon index, 1.42$_{-0.48}^{+0.52}$, consistent with the value employed for other targets.  The spectrum also shows excess consistent with Fe K$\beta$ emission. In order to better represent the spectrum below 2\,keV, we added a second MEKAL component with fixed 0.08\,keV temperature.

The spectrum of NGC\,1194 is still richer,  exhibiting evidence above 5\,keV for direct transmission from the central engine, which we modeled as an absorbed power law with  N$_{\rm H}=1.06_{-0.24}^{+0.36} \times 10^{24}$ cm$^{-2}$ and $\Gamma=2.58_{-0.44}^{+0.64}$.  The column presumably arises in the reflector.  About 70\% of the estimated apparent source luminosity (2-10\,keV) arises from transmission.  Assuming that this component corresponds to  emission from the central engine, the unabsorbed luminosity is 4.4$\times$10$^{42}$ erg s$^{-1}$, and the inferred  $\Omega/2\pi$ for the reflector is on the order 0.1, which is consistent with that inferred for  cold gas in, e.g., NGC 1068 \citep{col02,pounds06}, Mrk 3 \citep{pounds05} , NGC 4945 \citep{done03}.

The spectrum of NGC\,1194 also includes what appear to be K$\alpha$ lines from Si, S, and Ca (Tab.\,2).  The profiles are broadened, which we hypothesize is due to photoionization and blending of emission from different states, as observed in IC\,2560 \citep{mad06}.  Because our primary intent is characterization of absorption columns, we only note in passing that treatment of photoionization could be done in detail using an XSTAR model \citep[e.g.,][]{bk01}. 

\section{Discussion }

We have combined published estimates and limits on N$_{\rm H}$ with results of our analyses to obtain a sample of 42 AGN (out of 104 known maser hosts).  The sample includes: (1) 31 AGN that exhibit megamaser and weaker emission \citep{zha06}, (2) two AGN listed by  \citet{zha06} but without known columns  -- NGC\,591, Mrk\,1, (3) five AGN treated here -- Tab.\,2, and (4) four nuclei in which maser emission has been recently discovered -- NGC\,17 \citep{kon07}; NGC\,3081, NGC\,4253, and NGC\,3783 (Braatz \etal, in preparation).  
Columns are taken from Zhang \etal -- 31 objects, except NGC\,5256 \citep[cf.][]{gua05}; this work -- 5 objects; \citet{gua05} -- Mrk\,1, NGC\,17, and NGC\,591; \cite{ree04} -- NGC\,3783; \citet{mai98} -- NGC\,3081; and \cite{pounds03} --  NGC\,4253. In order of precedence, we adopt results from SAX, XMM, Chandra and ASCA, except for NGC\,3079, where we judged the XMM lower limit \citep{cappi06} to be more consistent with SAX data  than larger limits based on SAX data alone \citep{iyo01}.

\begin{figure}
\begin{center}
\scalebox{0.9}[0.85]{\plotone{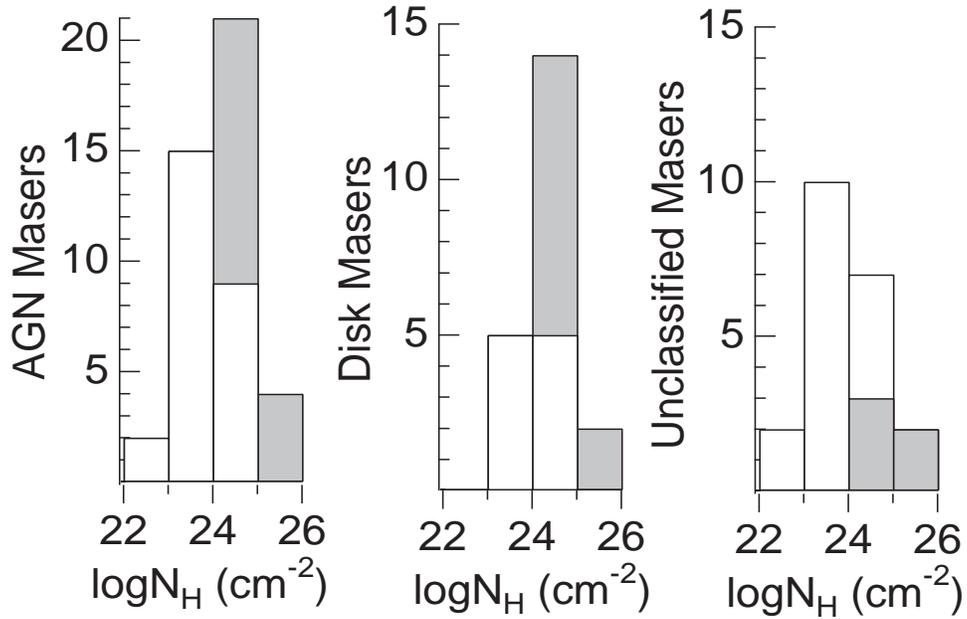}}
\caption{({\it left})  Distribution of N$_{\rm H}$ in AGN that host masers and where X-ray data are available.  Shaded bars indicate lower limits. ({\it middle}) Distribution for disk masers. Based on geometry, disk maser hosts are plausibly anticipated to constitute a more homogeneous sample with respect to N$_{\rm H}$. ({\it right}) Distribution for maser AGN that are not identifiable as disk masers at this time.}
\label{fig2}
\end{center}
\end{figure}

Of the 42 AGN, 60\% (25) are Compton thick (Tab.\,\ref{tabnh}; Fig.\,\ref{fig2}).  This fraction is consistent with that obtained for the smaller sample of \citet{zha06}, 58\% (18 of 31).
Within the current sample of 42, a greater fraction of  recognized disk-maser galaxies are Compton thick, 76\% (16 of 21).  The distribution of columns for disk masers is significantly skewed and non-Gaussian (Fig.\,\ref{fig2}).  Substitution (over time) of N$_{\rm H}$ estimates for lower limits will exagerrate this.  In contrast, the distribution for AGN masers without evidence of origins in  edge-on disks includes a larger proportion of Compton-thin sources.  A Kolmogorov-Smirnov (KS) test indicates a $< 3.2\%$ probability that  the two distributions are drawn from a single parent. To account for lower limits in our test, we constructed a Monte Carlo simulation of 1000 trials, sampled a uniformly distributed random variable of $10^{24}$ and $10^{26}$\,cm$^{-2}$ for each galaxy with a limit, and compiled a distribution of KS statistics.   Boundaries of $10^{25.5}$ and  $10^{28}$ yielded probabilities of $\la3\%$ as well.  

Distinctions between N$_{\rm H}$ distributions for AGN that host acknowledged disk masers and those that do not may be still greater than estimated. Maser classifications depend on identification of spectroscopic markers: highly red and blue-shifted emission bracketing the systemic velocity.  Disks may be oriented close to edge-on, resulting in large columns toward the central engine, while these spectroscopic markers may nonetheless be absent (e.g., high-velocity masers may be beamed away from the observer due  to warping; cf. Miyoshi \etal\ 1995).

\begin{deluxetable}{lccc}
\tablecolumns{4}
\tablewidth{3in}
\tablewidth{4.4in}
\tablecaption{Column Measurements Among Known Maser AGN}
\tablehead{
\colhead{Sample} & 
\colhead{$>10^{23}$ cm$^{-2}$} & 
\colhead{$>10^{24}$ cm$^{-2}$} &
\colhead{No.} 
} 
\startdata
Disk masers\tablenotemark{(a)}   & 100\%  & 76\%  & 21\\
All AGN masers                                & 95\% & 60\% & 42 \\
Megamasers\tablenotemark{(b)}  & 87\%  & 58\%  & 31 \\
\vspace{-0.08in}
\enddata
\tablenotetext{(a)}{In addition to five sources in Tab.\,2, we recognize 16 other reported disk-maser AGN with accompanying estimates of N$_{\rm H}$ or limits: NGC\,591, NGC\,1068, NGC\,1386, NGC\,2273, NGC\,3079, NGC\,3393, NGC\,4051, NGC\,4258, NGC\,4388, NGC\,4945, NGC\,5728, IC\,2560, Mrk\,1, Mrk\,1210, \hbox{Circinus, and 3C\,403.} }
\tablenotetext{(b)}{We include 26 megamasers with estimates of N$_{\rm H}$ or limits as listed by \citet{zha06}, four objects in Tab. 2 (NGC\,1194, NGC\,6926, Mrk\,34, Mrk\,1419), and Mrk\,1. }  
\label{tabnh}
\end{deluxetable}

\citet{zha06} discussed an alternate sub-sample, selected based on apparent maser luminosity (i.e., megamasers).  Within the current sample of 42 AGN with masers, we count 58\% (18 of 31) of recognized megamaser hosts are Compton thick (13 from \citet{zha06}, plus NGC\,1194, NGC\,6926, Mrk\,1, Mrk\,1419, and Mrk\,34), matching the incidence among maser AGN broadly (Tab.\,\ref{tabnh}).  We link the larger incidence of columns $>10^{24}$\,cm$^{-2}$ among disk-masers to a simple model geometry (edge-on orientation) and contrast this with selection based on maser luminosity, which is nearly always estimated with the convenient but inaccurate assumption of isotropic emission, without knowledge of beam angles and orientation to the line of sight.
 
Columns $\ga 10^{23}$ cm$^{-2}$ are more  common in maser AGN than in a sample of 22 hard X-ray (14-195\,keV) selected, similarly nearby AGN observed with SWIFT  \citep{win08}, which includes no maser hosts (95\% vs 50\%; Tab.\,\ref{tabnh}). However, for N$_{\rm H}>10^{24}$\,cm$^{-2}$, incidence rates are similar (60\% vs 50\%).  
For masers in all classes of AGN, we estimate a statistically similar frequency of Compton-thick columns to that reported by \citet{gua05} and \citet{ris99} among Seyfert\,2 objects: $46\%$ (22 of 48) and 65\% (22 of 34), respectively.  However, this is somewhat below the rate for masers associated with  Seyfert\,2 nuclei, 74\% (20 of 27).  The difference is suggestive but comparison of column distributions of maser systems and local type-2 AGN is complicated by sample incompleteness in luminosity and distance, overlap between the X-ray and maser samples (e.g., 15/45 sources for Risaliti \etal\ 1999), and uncertainty in X-ray luminosities for obscured AGN (though aperture and extinction-corrected [OIII] luminosities could be used in place of X-ray estimates, e.g., Mel\'endez \etal\ 2008).  

Demonstrated correlation between incidence of maser emission and high N$_{\rm H}$ may suggest a new strategy for future efforts to identify obscured AGN.  This is of particular interest if such objects are indeed substantial contributors to the Hard X-ray Background.  Heavily obscured  AGN typically have relatively low apparent luminosities and require long  integrations  (e.g., $> 100$ ksec  to obtain $>10^3$ counts for $cz=10^3-10^4$\,km\,s$^{-1}$).  In contrast, maser emission in AGN as distant as $cz\sim10^4$\,km\,s$^{-1}$  can be detected  with integrations on the order of 1\,ksec, using a well-instrumented 100m-dish antenna.  We characterize the relative efficiency of maser and pointed X-ray observations, $R$, by the ratio of Compton-thick AGN discoverable in a fixed time by each means, for a sample of optically identified type-2 objects.  We obtain $R\sim 3$ from $[10^5 f_m M_{CT}] / [10^3 X_{CT}]$, where $f_m$ is the maser detection rate  ($\sim 0.03$; Kondratko \etal\ 2006), $M_{CT}$ is the  Compton thick fraction of maser AGN ($\sim 0.6$; Tab.\,\ref{tabnh}), and $X_{CT}$ is the fraction of type-2 AGN found to be Compton thick in X-ray studies ($\sim 0.5$; e.g., Risaliti \etal\ 1999).

Deep, wide-field, hard X-ray surveys (e.g., NuSTAR) will extend synergy with maser studies, creating statistically significant X-ray and joint X-ray/radio samples.  Nonetheless, there will be practical constraints on survey fields ($5^\circ\times5^\circ$ for NuSTAR), and all-sky radio observations with well-defined completeness criteria could usefully guide pointed follow-up to enrich X-ray samples.

Strengthened correlation between disk-maser emission and large obscuring columns begs the question, is maser gas in disks responsible for obscuration in type-2 objects \citep{greenhill03,mad06}?  On the one hand, obscuration by cold disk gas at radii of 0.2-0.3 pc is inferred in NGC 4258 \citep{fru05, herrnstein05}, and for Circinus along other lines of sight \citep{greenhill03}.  On the other hand, time-variable N$_{\rm H}$ data indicate obscuration at radii of a few$\times0.01$ pc toward NGC\,1365, NGC\,4388, and NGC\,4151 \citep{elv04, ris07, puccetti07}.  Direct physical association of maser and obscuring gas volumes still appears to be case-specific.  Highly inclined structures that give rise to disk masers probably span a range of radii, at least extending inward with accretion flows.  Where these cross the line of sight to the central engine, and how extensively, likely depends on the detailed geometry of each system, requiring case study to establish.

\acknowledgments
We thank A. Fruscione for help with data reduction and M. Elvis, A. Siemiginowska and B. Wilkes for useful discussions. This research made extensive use of the NASA/IPAC Extragalactic Database and the NASA Astrophysical Data System Bibliographic Services. This work was supported in part by NASA grant NNG05GK24G and DoE contract to SLAC No. DE-AC3-76SF00515.

\end{document}